\documentclass[11pt]{article}
\usepackage{amsmath, amssymb, amsfonts, amsthm, mathtools}
\usepackage{graphicx}
\usepackage{color, dsfont}
\usepackage{graphicx}
\usepackage{mathabx}
\usepackage{multirow}
\usepackage{setspace}
\usepackage{tabularx}
\usepackage[labelsep=period]{caption}
\usepackage{subcaption}
\usepackage[hidelinks,bookmarks=false]{hyperref}
\usepackage[page,title,titletoc,header]{appendix}
\usepackage[top=1.5in, bottom=1.5in, left=1.5in, right=1.5in]{geometry}
\usepackage{url,indentfirst,comment,titlesec,secdot,enumerate,psfrag,epsf}
\usepackage[style=authoryear-comp,sortcites=false,firstinits=true, backend=bibtex,natbib=true,hyperref=true,maxbibnames=99,maxcitenames=2]{biblatex}
\usepackage{lipsum}
\usepackage{sectsty}

\titleformat*{\section}{\large\bfseries\centering}
\titleformat*{\subsection}{\normalsize\bfseries}	

\DeclareNameAlias{sortname}{last-first}

\newcolumntype{L}[1]{>{\raggedright\arraybackslash}p{#1}}
\newcolumntype{C}[1]{>{\centering\arraybackslash}p{#1}}
\newcolumntype{R}[1]{>{\raggedleft\arraybackslash}p{#1}}
\renewbibmacro{in:}{}
\sectionfont{\bfseries\large\raggedright}
\subsectionfont{\normalsize\raggedright}
\allowdisplaybreaks
\numberwithin{equation}{section}
\numberwithin{table}{section}
\numberwithin{figure}{section}

\author{Olanrewaju Akande, Jerome Reiter and Andr\'{e}s F. Barrientos \footnote{Olanrewaju M. Akande is PhD Candidate, Department of Statistical Science, Duke University, Durham, NC 27708 (E-mail: \href{mailto:olanrewaju.akande@duke.edu}{olanrewaju.akande@duke.edu}); Jerome P. Reiter is Professor of Statistical Science, Duke University, Durham, NC 27708 (E-mail: \href{mailto:jerry@stat.duke.edu}{jerry@stat.duke.edu}); and Andr\'{e}s F. Barrientos is Postdoctoral Associate, Department of Statistical Science, Duke University, Durham, NC 27708 (E-mail: \href{mailto:anfebar@stat.duke.edu}{anfebar@stat.duke.edu}). }  }
\title{Multiple Imputation of Missing Values in Household Data with Structural Zeros}
\date{}

\begin{document}
\maketitle

\begin{abstract}
\noindent 
We present an approach for imputation of missing items in multivariate categorical data nested within households.  The approach relies on a latent class model that (i) allows for household-level and individual-level variables, (ii) ensures that impossible household configurations have zero probability in the model, and (iii) can preserve multivariate distributions both within households and across households.  We present a Gibbs sampler for estimating the model and generating imputations.  We also describe strategies for improving the computational efficiency of the model estimation.  We illustrate the performance of the approach with data that mimic the variables collected in typical population censuses. 
\end{abstract}


\section{Introduction}
In many population censuses and demographic surveys, statistical agencies collect data on individuals grouped within houses. In the U.\ S.\ decennial census, for example, the Census Bureau collects the age, race, sex, and relationship to the household head for every individual in the household, as well as whether or not the residents own the house.  After collection, agencies share these datasets for secondary analysis, either as tabular summaries, public use microdata samples, or restricted access files.

When creating these data products, agencies typically have to deal with item nonresponse both for individual-level variables and household-level variables.  They typically do so using some type of imputation procedure.  Ideally, these procedures satisfy three desiderata.  First, the imputations preserve the joint distribution of the variables as best as possible.  As part of this, the procedure should preserve relationships within households.  For example, the missing race of a spouse likely, but certainly not definitely, matches the race of the household head; the imputation procedure should reflect that. Second, the imputations respect structural zeros.  For example, a daughter's age cannot exceed her biological mother's age.  The imputations should not create impossible combinations of individuals in the same household. Third, the imputation procedure allows for appropriate uncertainty to be propagated in subsequent analyses of the data.  

Typical approaches to imputation of missing household items use some variant of hot deck imputation \citep{KaltonKasprzyk1986, AndridgeLittle2010}. However, depending on how the hot deck is implemented, it may not satisfy one or more of the desiderata.  Indeed, we are not aware of any hot deck imputation procedure for household data that satisfies all three explicitly. 
An alternative is to estimate a model that describes the joint distribution of all the variables, and impute missing values from the implied predictive distributions in the model. 
For household data, one such model is the nested data Dirichlet process mixture of products of multinomial distributions (NDPMPM) model of \citet{HuEtAl2018}, which assumes that (i) each household is a member of a household-level latent class, and (ii) each individual is a member of an individual-level latent class nested within its household-level latent class.  The model 
assigns zero probability to combinations corresponding to structural zeros, and also handles both household-level and individual-level variables simultaneously. 
The NDPMPM is appealing as an imputation engine, as it can preserve multivariate associations while avoiding imputations that result in impossible households.
The NDPMPM is related to  models proposed by \citet{Vermunt2003,Vermunt2008} and \citet{BenninkEtAl2016}, although these are used for regression rather than multivariate imputation and do not deal with structural zeros.

\citet{HuEtAl2018} use the NDPMPM to generate synthetic datasets \citep{Rubin1993, RaghunathanRubin2001, ReiterRaghunathan2007} for statistical disclosure limitation, but they do not describe how to use it for imputation of missing data. 
We do so in this article.  With structural zeros in the NDPMPM, the conditional distributions of the missing values given the observed values are not available in closed form.  
We therefore add a rejection sampling step to the Gibbs sampler used by \citet{HuEtAl2018}, which generates completed datasets as byproducts of the Markov chain Monte Carlo (MCMC) algorithms used to estimate the model. 
These completed datasets can be analyzed using multiple imputation inferences \citep{Rubin1987}.  We also present two new strategies for speeding up the computations with NDPMPMs, namely (i) turning data for the household head into household-level variables rather than individual-level variables, and (ii) using an approximation to the likelihood function. These  scalable innovations are necessary, as the NDPMPM is computationally quite intensive even without missing data.  The speed-up strategies also can be employed when using the NDPMPM to generate synthetic data.

The remainder of this article is organized as follows. In Section \ref{NDPMPM}, we review the NDPMPM model in the presence of structural zeros and the MCMC sampler for fitting the model without missing data. In Section \ref{MissingData}, we extend the MCMC sampler for the NDPMPM model to allow for missing data.  In Section \ref{SpeedUp}, we present the two strategies for speeding up the MCMC sampler. 
In Section \ref{simulations},  we present results of simulation studies used to examine the performance of the NDPMPM as a multiple imputation engine, using the two strategies for speeding up the run time. In Section \ref{Discussion}, we discuss findings, caveats and future work.

\section{Review of the NDPMPM Model} \label{NDPMPM}

\citet{HuEtAl2018} present the NDPMPM model including motivation for how it can preserve associations across variables and account for structural zeros.  Here, we summarize the model without detailed motivations, referring the reader to \citet{HuEtAl2018} for more information.    We begin with notation needed to understand the model and the Gibbs sampler, assuming complete data.  The presentation closely follows that in \citet{HuEtAl2018}.

\subsection{Notation and model specification}

Suppose the data contain $n$ households. Each household $i=1, \dots, n$ contains $n_i$ individuals, so that there are  $\sum_{i=1}^n n_i = N$ individuals in the data. Let $X_{ik} \in \{1, \ldots, d_k\}$ be the value of categorical variable $k$ for household $i$, which is assumed to be identical for all $n_i$ individuals in household $i$, where $k = p+1, \ldots, p+q$.  Let $X_{ijk} \in \{1, \ldots, d_k\}$ be the value of categorical variable $k$ for person $j$ in household $i$, where
$j = 1, \ldots, n_i$ and $k = 1, \ldots, p$.  Let $\textbf{X}_i = (X_{i(p+1)}, \dots, X_{i(p+q)}, X_{i11}, \dots, X_{in_ip})$  include all household-level and individual-level variables for the $n_i$ individuals in household $i$. 

Let $\mathcal{H}$ be the set of all household sizes that are possible in the population.  For all $h \in \mathcal{H}$, let $\mathcal{C}_h$ represent the set of all combinations of individual-level and household-level variables for households of size $h$, including impossible combinations; that is, $\mathcal{C}_h = \prod_{k=p+1}^{p+q} \{1, \ldots, d_k\} \prod_{j=1}^{h} \prod_{k=1}^{p} \{1, \ldots, d_k\}$.  Let $\mathcal{S}_h \subset \mathcal{C}_h$ represent the set of impossible combinations, i.e., those that are structural zeros, for households of size $h$. These include combinations of variables within any individual, e.g., a three year old person cannot be a spouse, or across individuals in the same household, e.g., a person cannot be older than his biological parents. 
Let $\mathcal{C} = \bigcup_{h \in \mathcal{H}} \mathcal{C}_h$ and $\mathcal{S} = \bigcup_{h \in \mathcal{H}} \mathcal{S}_h$.

Although the NDPMPM model we use restricts the support of $\textbf{X}_i$ to $\mathcal{C} - \mathcal{S}$, it is helpful for understanding the model to begin with no restrictions on the support
of $\textbf{X}_i$.
Each household $i$ belongs to one of $F$ classes representing latent household types.  For $i=1, \dots, n$, let $G_i \in \{1, \dots, F\}$ indicate the household class for household $i$.
Let $\pi_g = \Pr(G_i = g)$ be the probability that household $i$ belongs to class $g$.
Within any class, all household-level variables follow independent, multinomial distributions.
For any $k \in \{p+1,\ldots,p+q\}$ and any $c \in \{1,\ldots,d_k\}$, let $\lambda_{gc}^{(k)} = \Pr(X_{ik} = c | G_i = g)$ for any class $g$, where $\lambda_{gc}^{(k)}$ is the same value for every household in class $g$.  Let $\pi = \{\pi_1, \ldots \pi_F\}$, and $\lambda = \{\lambda^{(k)}_{gc}: c = 1, \ldots, d_k; k = p+1, \ldots, p+q; g = 1, \ldots, F\}$.

Within each household class, each individual belongs to one of $S$ individual-level latent classes. 
For $i=1, \dots, n$ and $j=1, \dots, n_i$, let $M_{ij}$ represent the individual-level latent class of individual $j$ in household $i$. Let $\omega_{gm} = \Pr(M_{ij} = m | G_i = g)$ be the probability that individual $j$ in household $i$ belongs to individual-level class $m$ nested within household-level class $g$. 
Within any individual-level class, all individual-level variables follow independent, multinomial distributions.
For any $k \in \{1,\ldots,p\}$ and any $c \in \{1,\ldots,d_k\}$, let $\phi_{gmc}^{(k)} = \Pr(X_{ijk} = c | (G_i, M_{ij}) = (g,m))$ for the class pair $(g,m)$, where $\phi_{gmc}^{(k)}$ is the same value for every individual in the class pair $(g,m)$. Let $\omega = \{\omega_{gm}: g = 1, \ldots, F; m=1, \ldots, S \}$, and $\phi = \{\phi^{(k)}_{gmc}: c = 1, \ldots, d_k; k = 1, \ldots, p; m=1, \ldots, S ; g = 1, \ldots, F\}$.

For purposes of the Gibbs sampler in Section \ref{NDPMPMSampler}, it is useful to distinguish values of $\mathbf{X}_i$ that satisfy all the structural zero constraints from those that do not.  Let the superscript ``$1$'' indicate that a random variable has support only on $\mathcal{C} - \mathcal{S}$. For example, $\textbf{X}_i^1$ represents data for a household with values restricted only on $\mathcal{C} - \mathcal{S}$, i.e., not an impossible household, whereas $\textbf{X}_{i}$ represents data for a household with any values in $\mathcal{C}$. Let $\mathcal{X}^1$ be the observed data comprising $n$ households, that is, a realization of $(\textbf{X}^1_{1}, \ldots, \textbf{X}^1_{n})$. The kernel of the NDPMPM, $\Pr(\mathcal{X}^1 | \theta)$, is
\begin{equation} \label{StrucZeroLikelihood}
\textrm{L}(\mathcal{X}^1 | \theta) = \prod_{i=1}^n \sum_{h \in \mathcal{H}} \mathds{1}\{n_i = h \} \mathds{1}\{\textbf{X}_i^1 \notin \mathcal{S}_h \} \left[ \sum_{g=1}^F \pi_g  \prod^{p+q}_{k = p+1} \lambda^{(k)}_{gX^1_{ik}} \prod^{h}_{j=1} \sum_{m=1}^S \omega_{gm}\prod^p_{k = 1} \phi^{(k)}_{gmX^1_{ijk}} \right],
\end{equation}
where $\theta$ includes all the parameters, and $\mathds{1}\{.\}$ equals one when the condition inside the $\{\}$ is true and equals zero otherwise.

For all $h \in \mathcal{H}$, let $n_{1h} = \sum_{i=1}^n \mathds{1}\{n_i = h\}$ be the number of households of size $h$ in $\mathcal{X}^1$ and $\pi_{0h}(\theta) = \Pr(\textbf{X}_i \in \mathcal{S}_h | \theta) $
As stated in \citet{HuEtAl2018}, the normalizing constant in the likelihood in (\ref{StrucZeroLikelihood}) is $\prod_{h \in \mathcal{H}}(1 - \pi_{0h}(\theta))^{n_{1h}} $. Therefore, the posterior distribution is
\begin{equation} \label{NDPMPMTruncatedPosterior}
\Pr(\theta | \mathcal{X}^1, T(\mathcal{S})) \propto \Pr(\mathcal{X}^1 | \theta) \Pr(\theta) = \dfrac{1}{\prod_{h \in \mathcal{H}}(1 - \pi_{0h}(\theta))^{n_{1h}} } \textrm{L}(\mathcal{X}^1 | \theta) \Pr(\theta)
\end{equation}
where $T(\mathcal{S})$ 
emphasizes that the density is for the  NDPMPM with support restricted to  $\mathcal{C} - \mathcal{S}$.

The likelihood in \eqref{StrucZeroLikelihood} can be written as a generative model of the form
\begin{align}
\phantom{X_{ijk} | G_i, M_{ij}, \phi, n_i}
&\begin{aligned} \label{ModelSpecification1}
\mathllap{X_{ik} | G_i, \lambda} & \sim \textrm{Discrete}(\lambda^{(k)}_{G_i1}, \ldots, \lambda^{(k)}_{G_id_k}) \\
&\qquad \forall i = 1, \ldots, n \ \textrm{and} \  k = p + 1, \ldots, p + q
\end{aligned}\\
&\begin{aligned} \label{ModelSpecification2}
\mathllap{X_{ijk} | G_i, M_{ij}, \phi, n_i}  & \sim \textrm{Discrete}(\phi^{(k)}_{G_iM_{ij}1}, \ldots, \phi^{(k)}_{G_iM_{ij}d_k}) \\
&\qquad \forall i = 1, \ldots, n \ , \ j = 1, \ldots, n_i \ \textrm{and} \ k = 1, \ldots, p
\end{aligned}\\
&\begin{aligned} \label{ModelSpecification3}
\mathllap{G_i | \pi} & \sim \textrm{Discrete}(\pi_1, \ldots, \pi_F) \\
&\qquad \forall i = 1, \ldots, n
\end{aligned}\\
&\begin{aligned} \label{ModelSpecification4}
\mathllap{M_{ij} | G_i, \omega, n_i} & \sim \textrm{Discrete}(\omega_{G_i1}, \ldots, \omega_{G_iS}) \\
&\qquad \forall i = 1, \ldots, n \ \textrm{and} \  j = 1, \ldots, n_i
\end{aligned}
\end{align}
where the Discrete distribution refers to the multinomial distribution with sample size equal to one. We restrict the support of each $\mathbf{X}_i$ to ensure the model assigns zero probability to all combinations in $\mathcal{S}$ as desired. The model in (\ref{ModelSpecification1}) to (\ref{ModelSpecification4}) can be used without restricting the support to  $\mathcal{C} - \mathcal{S}$.  This ignores all structural zeros.  While not appropriate for the joint distribution of household data, this model turns out to useful for the Gibbs sampler. We refer to the generative model in (\ref{ModelSpecification1}) to (\ref{ModelSpecification4}) with support on all of $\mathcal{C}$ as the untruncated NDPMPM.  For contrast, we call the model in \eqref{StrucZeroLikelihood} the truncated NDPMPM.

For prior distributions, we follow the recommendations of \citet{HuEtAl2018}. We use independent uniform Dirichlet distributions as priors for $\lambda$ and $\phi$, and the truncated stick-breaking representation of the Dirichlet process as priors for $\pi$ and $\omega$ \citep{Sethuraman1994, DunsonXing2009, SiReiter2013, Manrique-VallierReiter2014b},
\begin{align}
\phantom{\omega_{gm} }
&\begin{aligned} \label{DirichletPriorI}
\mathllap{\lambda_g^{(k)} } & = (\lambda^{(k)}_{g1}, \ldots, \lambda^{(k)}_{gd_k}) \sim \textrm{Dirichlet}(1,\ldots, 1) \\ 
\end{aligned}\\
&\begin{aligned} \label{DirichletPriorII}
\mathllap{\phi_{gm}^{(k)} } & = (\phi^{(k)}_{gm1}, \ldots, \phi^{(k)}_{gmd_k}) \sim \textrm{Dirichlet}(1,\ldots, 1) \\ 
\end{aligned}\\
&\begin{aligned}
\mathllap{\pi_g } & = u_g \prod_{f < g} (1 -  u_f) \ \textrm{for} \ g = 1, \ldots F \\
\end{aligned}\\
&\begin{aligned}
\mathllap{u_g } & \sim \textrm{Beta}(1,\alpha) \ \textrm{for} \ g = 1, \ldots, F-1, \ u_F = 1  \\
\end{aligned}\\
&\begin{aligned} \label{GammaPriorI}
\mathllap{\alpha } & \sim \textrm{Gamma}(0.25, 0.25) \\ 
\end{aligned}\\
&\begin{aligned}
\mathllap{\omega_{gm} } & = v_{gm} \prod_{s < m} (1 -  v_{gs}) \ \textrm{for} \ m = 1, \ldots S \\
\end{aligned}\\
&\begin{aligned}
\mathllap{v_{gm} } & \sim \textrm{Beta}(1,\beta_g) \ \textrm{for} \ m = 1, \ldots, S-1, \ v_{gS} = 1  \\
\end{aligned}\\
&\begin{aligned} \label{GammaPriorII}
\mathllap{\beta_g } & \sim \textrm{Gamma}(0.25, 0.25). \\ 
\end{aligned}
\end{align}

We set the parameters for the Dirichlet distributions in (\ref{DirichletPriorI}) and (\ref{DirichletPriorII}) to $\mathbf{1}_{d_k}$ (a $d_k$-dimensional vector of ones) and the parameters for the Gamma distributions in (\ref{GammaPriorI}) and (\ref{GammaPriorII}) to $0.25$ to represent vague prior specifications. We also set $\beta_g = \beta$ for computational expedience. For further discussion on prior specifications, see \citet{HuEtAl2018}.

Conceptually, the latent household-level classes can be interpreted as clusters of households with similar compositions, e.g., households with children or households in which no one is related. Similarly, the latent individual-level classes can be interpreted as clusters of individuals with similar characteristics, e.g., older male spouses or young female children.  However, for purposes of imputation, we do not care much about interpreting the classes, as they serve mainly to induce dependence across variables and individuals in the joint distribution.

It is important to select $F$ and $S$ to be large enough to ensure accurate estimation of the joint distribution.  However, we also do not want to make $F$ and $S$ so large as to produce many empty classes in the model estimation.  Allowing many empty classes increases computational running time without any corresponding increase in estimation accuracy.  This can be especially problematic in the Gibbs sampler for the truncated NDPMPM, as these empty classes can introduce mass in regions of the space where impossible combinations are likely to be generated.  This slows down the convergence of the Gibbs sampler.  

We therefore recommend following the strategy in \citet{HuEtAl2018} when setting $(F, S)$.  Analysts can start with moderate values for both, say between 10 and 15, in initial tuning runs. After convergence, analysts examine posterior samples of the latent classes to check how many individual-level and household-level latent classes are occupied. Such posterior predictive checks can provide evidence for the case that larger values for $F$ and $S$ are needed. If the numbers of occupied household-level classes hits $F$, we suggest increasing $F$. If the number of occupied individual-level classes hits $S$, we suggest increasing $F$ first but then increasing $S$, possibly in addition to $F$, if increasing $F$ alone does not suffice. 
When posterior predictive checks do not provide evidence that larger values of $F$ and $S$ are needed, analysts need not increase the number of classes, as doing so is not expected to improve the accuracy of the estimation.  We note that similar logic is used in other mixture model contexts \citep{Walker2007, SiReiter2013, Manrique-VallierReiter2014b,MurrayReiter2016}.

\subsection{MCMC sampler for the NDPMPM} \label{NDPMPMSampler}

\citet{HuEtAl2018} use a data augmentation strategy \citep{Manrique-VallierReiter2014b} to estimate the posterior distribution in (\ref{NDPMPMTruncatedPosterior}). They assume that the observed data $\mathcal{X}^1$, which includes only feasible households, is a subset from a hypothetical sample $\mathcal{X}$ of $(n + n_0)$ households directly generated from the untruncated NDPMPM. That is, $\mathcal{X}$ is generated on the support $\mathcal{C}$ where all combinations are possible and structural zeros rules are not enforced, but we only observe the sample of $n$ households $\mathcal{X}^1$ that satisfy the structural zero rules and do not observe the sample of $n_0$ households $\mathcal{X}^0 = \mathcal{X} - \mathcal{X}^1$ that fail the rules.

We use the strategy of \citet{HuEtAl2018} and augment the data as follows. For each $h \in \mathcal{H}$, 
we simulate $\mathcal{X}$ from the untruncated NDPMPM, stopping when the number of simulated feasible households in $\mathcal{X}$ directly matches $n_{1h}$ for all $h \in \mathcal{H}$. We replace the simulated feasible households in $\mathcal{X}$ with $\mathcal{X}^1$, thus, assuming that $\mathcal{X}$ already contains $\mathcal{X}^1$ and we only need to generate the part $\mathcal{X}^0$ that fall in $\mathcal{S}$. Given a draw of $\mathcal{X}$, we draw $\theta$ from posterior distribution defined by the untruncated NDPMPM, treating $\mathcal{X}$ as the observed data. This posterior distribution can be estimated using a blocked Gibbs sampler \citep{IshwaranJames2001, SiReiter2013}.

We now present the full MCMC sampler for fitting the truncated NDPMPM. Let $\textbf{G}^0$ and $\textbf{M}^0$ be vectors of the latent class membership indicators for the households in $\mathcal{X}^0$ and $n_{0h}$ be the number of households of size $h$ in $\mathcal{X}^0$, with $n_0 = \sum_h n_{0h}$. In each full conditional, let ``--'' represent conditioning on all other variables and parameters in the model. At each MCMC iteration, we do the following steps.

\begin{enumerate}
	\item[S1.] Set $\mathcal{X}^0 = \textbf{G}^0 = \textbf{M}^0 = \emptyset$. For each $h \in \mathcal{H}$, repeat the following:
	
	\begin{enumerate}
		\item Set $t_0 = 0$ and $t_1 = 0$.
		
		\item Sample $G_i^0 \in \{1, \ldots, F \} \sim \textrm{Discrete}(\pi_1^{\star\star}, \ldots, \pi_F^{\star\star})$ where $\pi_g^{\star\star} \  \propto \ \lambda^{(k)}_{gh} \pi_g $ and $k$ is the index for the household-level variable ``household size''. 
		
		\item For $j = 1, \ldots, h$, sample $M^0_{ij} \in \{1, \ldots, S\} \sim \textrm{Discrete}(\omega_{G^0_i1}, \ldots, \omega_{G^0_iS})$.
		
		\item Set $X^0_{ik} = h$, where $X^0_{ik}$ corresponds to the variable for household size. Sample the remaining household-level and individual-level values using the likelihoods in (\ref{ModelSpecification1}) and (\ref{ModelSpecification2}). Set the household's simulated value to $\textbf{X}^0_i$.
		
		\item If $\textbf{X}^0_i \in \mathcal{S}_h$, let $t_0 = t_0 + 1$, $\mathcal{X}^0 = \mathcal{X}^0 \cup \textbf{X}^0_i$, $\textbf{G}^0 = \textbf{G}^0 \cup G^0_i$ and $\textbf{M}^0 = \textbf{M}^0 \cup \{M_{i1}^0, \ldots, M_{ih}^0 \}$. Otherwise set $t_1 = t_1 + 1$. \label{StepF}
		
		\item If $t_1 < n_{1h}$, return to step (b). Otherwise, set $n_{0h} = t_0$. \label{StepG}
	\end{enumerate}
	
	\item[S2.] For observations in $\mathcal{X}^1$,
	
	\begin{enumerate}
		\item Sample $G_i \in \{1, \ldots, F \} \sim \textrm{Discrete}(\pi_1^\star, \ldots, \pi_F^\star)$ for $i = 1, \ldots, n$, where 
		$$\pi_g^\star = \Pr(G_i = g | - ) = \dfrac{\pi_g \left[\prod\limits^q_{k=p+1} \lambda^{(k)}_{gX^1_{ik}} \left(\prod\limits^{n_i}_{j=1} \sum\limits^S_{m=1}\omega_{gm}\prod\limits^p_{k=1} \phi^{(k)}_{gmX^1_{ijk}} \right) \right] }{\sum\limits^F_{f=1} \pi_f \left[\prod\limits^q_{k=p+1} \lambda^{(k)}_{fX^1_{ik}} \left(\prod\limits^{n_i}_{j=1} \sum\limits^S_{m=1}\omega_{gm}\prod\limits^p_{k=1} \phi^{(k)}_{fmX^1_{ijk}} \right) \right] }$$
		for $g = 1, \ldots, F$. Set $G_i^1 = G_i$.
		
		\item Sample $M_{ij} \in \{1, \ldots, S\} \sim \textrm{Discrete}(\omega_{G_i^11}^\star, \ldots, \omega_{G_i^1S}^\star)$ for $i = 1, \ldots, n$ and $j = 1, \ldots, n_i$, where 
		$$\omega_{G_i^1m}^\star = \Pr(M_{ij} = m | - ) = \dfrac{\omega_{G_i^1m}\prod\limits^p_{k=1} \phi^{(k)}_{G_i^1mX^1_{ijk}}  }{ \sum\limits^S_{s=1}\omega_{G_i^1s}\prod\limits^p_{k=1} \phi^{(k)}_{G^1_isX^1_{ijk}}}$$
		for $m = 1, \ldots, S$. Set $M^1_{ij} = M_{ij}$
	\end{enumerate}
	
	\item[S3.] Set $u_F = 1$. Sample	
	\begin{equation*}
	\begin{split}
	u_g | - \ & \sim \textrm{Beta} \left(1 + U_g, \alpha + \sum^F_{f=g+1} U_f  \right), \ \  \pi_g = u_g \prod_{f<g} (1 - u_f) \\
	\textrm{where} \ \ U_g & = \sum^{n}_{i=1} \mathds{1}(G^1_i = g) + \sum\limits^{n_{0}}_{i=1} \mathds{1}(G_i^0 = g)
	\end{split}
	\end{equation*}
	for $g = 1, \ldots, F-1$.
	
	\item[S4.] Set $v_{gM} = 1$ for $g = 1, \ldots, F$. Sample
	\begin{equation*}
	\begin{split}
	v_{gm} | - \ & \sim \textrm{Beta} \left(1 + V_{gm}, \beta + \sum^S_{s=m+1} V_{gs}  \right), \ \ \omega_{gm} = v_{gm} \prod_{s<m} (1 - v_{gs}) \\
	\textrm{where} \ \ V_{gm} & = \sum^{n}_{i=1} \mathds{1}(M^1_{ij} = m, G^1_i = g) + \sum\limits^{n_{0}}_{i=1} \mathds{1}(M_{ij}^0 = m, G_i^0 = g)
	\end{split}
	\end{equation*}
	for $m = 1, \ldots, S-1$ and $g = 1, \ldots, F$.
	
	\item[S5.] Sample 
	\begin{equation*}
	\begin{split}
	\lambda_g^{(k)} | - & \sim \textrm{Dirichlet}\left(1 + \eta^{(k)}_{g1}, \ldots, 1 + \eta^{(k)}_{gd_k} \right) \\
	\textrm{where} \ \ \eta^{(k)}_{gc} & = \sum^{n}_{i|G_i^1 = g} \mathds{1}(X^1_{ik} = c) + \sum\limits^{n_{0}}_{i|G_i^0 = g} \mathds{1}(X_{ik}^0 = c)
	\end{split}
	\end{equation*}
	for $g = 1, \ldots, F$ and $k = p+1, \ldots, q$. 
	
	\item[S6.] Sample 
	\begin{equation*}
	\begin{split}
	\phi_{gm}^{(k)} | - & \sim \textrm{Dirichlet}\left(1 + \nu^{(k)}_{gm1}, \ldots, 1 + \nu^{(k)}_{gmd_k} \right) \\
	\textrm{where} \ \ \nu^{(k)}_{gmc} & = \sum^{n}_{i,j |\substack{G_i^1 = g, \\  M_{ij}^1 = m}} \mathds{1}(X^1_{ijk} = c) +  \sum\limits^{n_{0}}_{i,j |\substack{G_i^0 = g, \\  M_{ij}^0 = m}} \mathds{1}(X_{ijk}^0 = c)
	\end{split}
	\end{equation*}
	for $g = 1, \ldots, F$, $m = 1, \ldots, S$ and $k = 1, \ldots, p$.
	
	\item[S7.] Sample
	$$ \alpha  | - \sim \textrm{Gamma}\left(a_\alpha + F - 1, b_\alpha - \sum^{F-1}_{g=1} \textrm{log}(1-u_g) \right).$$
	
	\item[S8.] Sample
	$$ \beta  | - \sim \textrm{Gamma}\left(a_\beta + F \times (S - 1), b_\beta - \sum^{S-1}_{m=1} \sum^{F}_{g=1} \textrm{log}(1-v_{gm}) \right).$$
\end{enumerate}

This Gibbs sampler is implemented in the R software package ``NestedCategBayesImpute'' \citep{WangEtAl2016}.  The software can  be used to generate synthetic versions of the original data, but it requires all data to be complete.

\section{Handling Missing Data Using the NDPMPM} \label{MissingData}

We modify the Gibbs sampler for the truncated NDPMPM to incorporate missing data. 
For $i = 1, \ldots, n$, let $\textbf{a}_i = (a_{i(p+1)}, \ldots, a_{i(p+q)})$ be a vector with $a_{ik} = 1$ when household-level variable $k \in \{p+1, \ldots, p+q\}$ in $\textbf{X}_i^1$ is missing, and $a_{ik} = 0$ otherwise.  For $i = 1, \dots, n$ and $j = 1, \dots, n_i$, let $\textbf{b}_{ij} = (b_{ij1}, \ldots, b_{ijp})$ be a vector with $b_{ijk} = 1$ when individual-level variable $k \in \{1, \ldots, p\}$ for individual $j \in \{1, \ldots, n_i\}$ in $\textbf{X}_i^1$ is missing, and $b_{ijk} = 0$ otherwise.
For each household $i$, let $\textbf{X}_i^1 = (\textbf{X}_i^{\textrm{obs}},\textbf{X}_i^{\textrm{mis}})$, where $\textbf{X}_i^{\textrm{obs}}$ comprise all data values corresponding to  $a_{ik} = 0$ and $b_{ijk} = 0$, and  $\textbf{X}_i^{\textrm{mis}}$ comprises all data values corresponding to  $a_{ik} = 1$ and $b_{ijk} = 1$.    
We assume that the data are missing at random \citep{Rubin1976}.  

To incorporate missing values in the Gibbs sampler, we need to sample from the full conditional of each variable in $\textbf{X}_i^{\textrm{mis}}$, conditioned on the variables for which $a_{ik} = 0$ and $b_{ijk} = 0$, at every iteration.
Thus, we add the ninth step,
\begin{enumerate}
	\item[S9.] For $i = 1, \ldots, n$, sample $\textbf{X}_i^{\textrm{mis}}$ from its full conditional distribution
	$$\Pr(\textbf{X}_i^{\textrm{mis}} | - ) \ \propto \ \mathds{1}\{\textbf{X}_i^1 \notin \mathcal{S}_h \} \ \left( \pi_{G^1_i} \prod\limits^{p+q}_{k | a_{ik} = 1} \lambda^{(k)}_{G^1_iX^1_{ik}} \prod\limits^{n_i}_{j=1} \omega_{G^1_iM^1_{ij}}\prod\limits^p_{k | b_{ijk} = 1} \phi^{(k)}_{G^1_iM^1_{ij}X^1_{ijk}} \right)$$
\end{enumerate}

Sampling from this conditional distribution is nontrivial because of the dependence among variables induced by the structural zero rules in each $\mathcal{S}_h$.   Because of the dependence, we cannot simply sample each variable independently using the likelihoods in (\ref{ModelSpecification1}) and (\ref{ModelSpecification2}). If we could generate the set of all possible completions for all households with missing entries, conditional on the observed values, then calculating the probability of each one and sampling from the set would be straightforward. Unfortunately, this approach is not practical when the size of each $\mathcal{S}_h$ is large. Even when the size of each $\mathcal{S}_h$ is modest, each household could have  different sets of completions, necessitating significant computing, storage, and memory requirements.

However, the full conditional in S9 takes a similar form as the kernel of the truncated NDPMPM in (\ref{StrucZeroLikelihood}), so that we can generate the desired samples through a second rejection sampling scheme. Essentially, we sample from an untruncated version of the full conditional $P^\star_{\textbf{X}_i^{\textrm{mis}}} = \pi_{G^1_i}\prod^{p+q}_{k | a_{ik} = 1} \lambda^{(k)}_{G^1_iX^1_{ik}} (\prod^{n_i}_{j=1} \omega_{G^1_iM^1_{ij}}\prod^p_{k | b_{ijk} = 1} \phi^{(k)}_{G^1_iM^1_{ij}X^1_{ijk}} ) $, until we obtain a valid sample that satisfies $\textbf{X}_i^1 \notin \mathcal{S}_h$; see the supplementary materials for a proof that this rejection sampling scheme results in a valid Gibbs sampler. Notice that since $P^\star_{\textbf{X}_i^{\textrm{mis}}}$ itself is untruncated, we can generate samples from it by sampling each variable independently using (\ref{ModelSpecification1}) and (\ref{ModelSpecification2}). We therefore replace step S9 with S9$^\prime$.
\begin{enumerate}
	\item[S9$^\prime$.] For $i = 1, \ldots, n$, sample $\textbf{X}_i^{\textrm{mis}}$ as follows.
	\begin{enumerate}
		\item For each missing household-level variable, that is, each variable where $k \in \{p+1, \ldots, p+q \}$ with $a_{ik} = 1$, sample $X_{ik}^{1}$ using (\ref{ModelSpecification1}). 
		
		\item For each missing individual-level variable, that is, each variable where $j = 1, \ldots, n_i$ and $k \in \{1, \ldots, p\}$ with $b_{ijk} = 1$, sample $X_{ijk}^{1}$ using (\ref{ModelSpecification2}).
		
		\item Set the sampled household-level and individual-level values to $\textbf{X}_i^{\textrm{mis}\star}$.
		
		\item Combine $\textbf{X}_i^{\textrm{mis}\star}$ with the observed $\textbf{X}_i^{\textrm{obs}}$, that is, set $\textbf{X}_i^{1\star} = (\textbf{X}_i^{\textrm{obs}},\textbf{X}_i^{\textrm{mis}\star})$. If $\textbf{X}_i^{1\star} \notin \mathcal{S}_h$, set $\textbf{X}_i^{\textrm{mis}} = \textbf{X}_i^{\textrm{mis}\star}$, otherwise, return to step (9$^\prime$a).
	\end{enumerate}
\end{enumerate}

To initialize each $\textbf{X}_i^{\textrm{mis}}$, we suggest sampling from the empirical marginal distribution of each variable $k$ using the available cases for each variable, and requiring that the household satisfies $\textbf{X}_i^1 \notin \mathcal{S}_h$.

\section{Strategies for Speeding Up the MCMC Sampler} \label{SpeedUp}

The rejection sampling step in the Gibbs sampler in Section \ref{NDPMPMSampler} can be inefficient when $\mathcal{S}$ is large \citep{Manrique-VallierReiter2014b,HuEtAl2018}, as the sampler tends to generate many impossible households before getting enough feasible ones.  
In addition, it takes computing time to check whether or not each sampled household satisfies all the structural zero rules. These computational costs are compounded when the sampler also incorporates missing values.  In this section, we present two strategies that can reduce the number of impossible households that the algorithm generates, thereby speeding up the sampler.  The supplementary material includes simulation studies showing that both strategies can speed up the MCMC significantly.

\subsection{Moving the household head to the household level} \label{SpeedUpMoveHH}

Many datasets include a variable recording the relationship of each individual to the household head.  There can be only one household head in any household. This restriction can account for a large proportion of the combinations in $\mathcal{S}$. As a simple working example, consider a dataset that contains $n=1000$ households of size two, resulting in a total of $N=2000$ individuals. Suppose the data contain no household-level variables and two individual-level variables, age and relationship to household head. Also, suppose age has 100 levels while relationship to household head has 13 levels, which include household head, spouse of the household head, etc. Then, $\mathcal{C}$ contains $13^2 \times 100^2 = 1.69 \times 10^6$ combinations.  Suppose the rule, ``each household must contain exactly one head,'' is the only structural zero rule defined on the dataset. Then, $\mathcal{S}$ contains $1.45 \times 10^6$ impossible combinations, approximately $86\%$ the size of $\mathcal{C}$.  If, for example, the model assigns uniform probability to all combinations in $\mathcal{C}$,
we would expect to sample about $(.86/.14) * 1000 \approx 6,143$ impossible households at every iteration to augment the $n$ feasible households. 

Instead, we treat the variables for the household head as a household-level characteristic.  This eliminates structural zero rules defined on the household head alone. 
Using the working example, moving the household head to the household level results in one new household-level variable, age of household head, which has 100 levels.  The relationship to household head variable can be ignored for  household heads. For others in the household, the relationship to household head variable now has 12 levels, with the level corresponding to ``household head'' removed. Thus, $\mathcal{C}$ contains  $12 \times 100^2 = 1.20 \times 10^5$ combinations, and $\mathcal{S}$ contains zero impossible combinations. 
We wouldn't even need to sample impossible households in the Gibbs sampler in Section \ref{NDPMPMSampler}.

In general, this strategy can reduce the size of $\mathcal{S}$ significantly, albeit usually not to zero as in the simple example here since $\mathcal{S}$ usually contains combinations resulting from other types of structural zero rules.  
This strategy is not a replacement for the rejection sampler in Section \ref{NDPMPMSampler}; rather, it is a data reformatting technique that can be combined with the sampler.

\subsection{Setting an upper bound on the number of impossible households to sample} \label{SpeedUpCapping}

To reduce computation time, we can put an upper bound on the number of sampled cases in  $\mathcal{X}^0$. One way to achieve this is to replace $n_{1h}$ in step S1(f) of Section \ref{NDPMPMSampler} with $\lceil n_{1h} \times \psi_h \rceil$, for some $\psi_h$ such that $1/\psi_h$ is a positive integer, so that we sample only approximately $\lceil n_{0h} \times \psi_h \rceil$ impossible households for each $h \in \mathcal{H}$. However, doing so underestimates the actual probability mass assigned to $\mathcal{S}$ by the model. We can illustrate this using the simple example of Section \ref{SpeedUpMoveHH}. Suppose the model assigns uniform probability to all combinations in $\mathcal{C}$ as before.  We set $\psi_2 = 0.5$, so that we sample approximately $3,072 = \lceil 6143 \times 0.5 \rceil$  impossible households in every iteration of the MCMC sampler. The probability of generating one impossible household is $3072 / (1000 + 3072) = 0.75$, a decrease from the actual value of 0.86. Therefore, we would underestimate the true contribution of $\{\mathcal{X}^0, \textbf{G}^0, \textbf{M}^0 \}$ to the likelihood.

To use the cap-and-weight approach, we need to apply a correction that re-weights the contribution of $\{\mathcal{X}^0, \textbf{G}^0, \textbf{M}^0 \}$ to the full joint likelihood. We do so using ideas akin to those used by \citet{ChambersSkinner2003, Savitskytoth2016}, approximating the likelihood of the full unobserved data with a ``pseudo'' likelihood using weights (the $1/\psi_h$'s). The impossible households only contribute to the full joint likelihood through the discrete distributions in (\ref{ModelSpecification1}) to (\ref{ModelSpecification4}). The sufficient statistics for estimating the parameters of the discrete distributions in (\ref{ModelSpecification1}) to (\ref{ModelSpecification4}) are the observed counts for the corresponding variables in the set $\{\mathcal{X}^1, \textbf{G}^1, \textbf{M}^1, \mathcal{X}^0, \textbf{G}^0, \textbf{M}^0 \}$, within each latent class for the household-level variables and within each latent class pair for the individual-level variables. Thus, for each $h \in \mathcal{H}$, we can re-weight the contribution of impossible households by multiplying the observed counts for households of size $h$ in $\{\mathcal{X}^0, \textbf{G}^0, \textbf{M}^0 \}$ by $1/\psi_h$ for the corresponding variable and latent classes. This raises the likelihood contribution of impossible households of size $h$ to the power of $1/\psi_h$. Clearly, $1/\psi_h$ need not be a positive integer. We require that only to make its multiplication with the observed counts free of decimals. We modify the Gibbs sampler to incorporate the cap-and-weight approach by replacing steps S1, S3, S4, S5 and S6; see the  supplementary materials for the modified steps.

Setting each $\psi_h = 1$ corresponds to the original rejection sampler, so that the two approaches should provide very similar results when $\psi_h$ near $1$.  Based on our experience, results of the cap-and-weight approach become significantly less accurate than the regular rejection sampler when $\psi_h < 1/4$.  The time gained using this speedup approach in comparison to the regular sampler depends on the features of the data and the specified values for the weights $\{\psi_h: h \in \mathcal{H} \}$.  To select the $\psi_h$'s, we suggest trying out different values---starting with values close to one---in initial runs of the MCMC sampler on a small random sample of the data. Analysts should examine the convergence and mixing behavior of the chains in comparison to the chain with all the $\psi_h$'s set to one, and select values that offer reasonable speedup while preserving convergence and mixing. This can be done quickly by comparing trace plots of a random set of parameters from the model that are not subject to label switching, such as $\alpha$ and $\beta$, or by examining marginal, bivariate and trivariate probabilities estimated from synthetic data generated from the MCMC.

\section{Empirical Study}\label{simulations}

To evaluate the performance of the NDPMPM as an imputation method, as well as the speed up strategies, we use data from the public use microdata files from the 2012 ACS, available for download from the United States Census Bureau (\url{http://www2.census.gov/acs2012_1yr/pums/}). We construct a population of 764,580 households of sizes $\mathcal{H} = \{2, 3, 4\}$, from which we sample $n=5,000$ households comprising  $N=13,181$ individuals.  We work with the variables described in Table \ref{MissingDataVariableDef}, which mimic those in the U.\ S.\ decennial census.  The structural zeros involve ages and relationships of individuals in the same house; see the supplementary material for a full list of rules that we used.  We move the household head to the household level as in Section \ref{SpeedUpMoveHH} to take advantage of the computational gains.  

\begin{table}[t!]
	\footnotesize
	\centering
	\caption{Description of variables used in the study. ``HH '' means household head.} \label{MissingDataVariableDef}
	\begin{tabular}[c]{p{0.4\textwidth}p{0.5\textwidth}}
		Description of variable & Categories \\ \hline 
		& \\
		\multicolumn{2}{l}{\underline{Household-level variables}} \\
		Ownership of dwelling & 1 = owned or being bought, 2 = rented\\
		Household size & 2 = 2 people, 3 = 3 people, 4 = 4 people \\
		Gender of HH & 1 = male, 2 = female \\
		Race of HH & 1 = white, 2 = black,\\
		& 3 = American Indian or Alaska native, \\
		& 4 = Chinese, 5 = Japanese, \\
		& 6 = other Asian/Pacific islander, 7 = other race, \\
		& 8 = two major races, \\
		& 9 = three or more major races \\
		Hispanic origin of HH & 1 = not Hispanic, 2 = Mexican, \\
		& 3 = Puerto Rican, 4 = Cuban, 5 = other \\
		Age of HH & 1 = less than one year old, 2 = 1 year old,\\
		& 3 = 2 years old, \ldots, 96 = 95 years old \\
		& \\
		\multicolumn{2}{l}{\underline{Individual-level variables}} \\
		Gender & same as ``Gender of HH'' \\
		Race & same as ``Race of HH'' \\
		Hispanic origin & same as ``Hispanic origin of HH'' \\
		Age & same as ``Age of HH'' \\
		Relationship to head of household & 1 = spouse, 2 = biological child, \\
		& 3 = adopted child, 4 = stepchild, 5 = sibling,\\
		& 6 = parent, 7 = grandchild, 8 = parent-in-law, \\
		& 9 = child-in-law, 10 = other relative, \\
		& 11 = boarder, roommate or partner, \\
		& 12 = other non-relative or foster child \\ \hline
	\end{tabular}
\end{table}

We introduce missing values using the following scenario. 
We let household size and age of household heads be fully observed. We randomly and independently blank 30\% of each variable for the remaining household-level variables. For individuals other than the household head, we randomly and independently blank 30\% of the values for gender, race and Hispanic origin.  We make age missing with rates 50\%, 20\%, 40\% and 30\% for values of the relationship variable in the sets \{2\}, \{3,4,5,10\}, \{7,9\} and \{6,8,11,12,13\}, respectively.  We make the relationship variable missing with rates 40\%, 25\%, 10\%, and 55\% for values of age in the sets  \{$x : x \leq 20$\}, \{$x : 20 < x \leq 50$\}, \{$x : 50 < x \leq 70$\}, and \{$x : x > 70$\}, respectively. This results in approximately 30\% missing values for both variables. About 8\% of the individuals in the sample are missing both the age and relationship variable, and 2\% are missing gender, age, and relationship jointly.  This mechanism results in data that technically are not missing at random, but we use the NDPMPM approach regardless to examine its potential in a complicated missingness mechanism.  Actual rates of item nonresponse in census data tend to be smaller than what we use here, but we use high rates to put the NDPMPM through a challenging stress test. We also introduce missing values using a missing completely at random scenario with rates in the 10\% range across all the variables. In short, the results are similar to those here, though more accurate due to the lower rates of missingness. See the supplementary material for the results.

 \begin{figure}[t]
	\centering
	\includegraphics[width=\textwidth, height = 3.1in, angle=0]{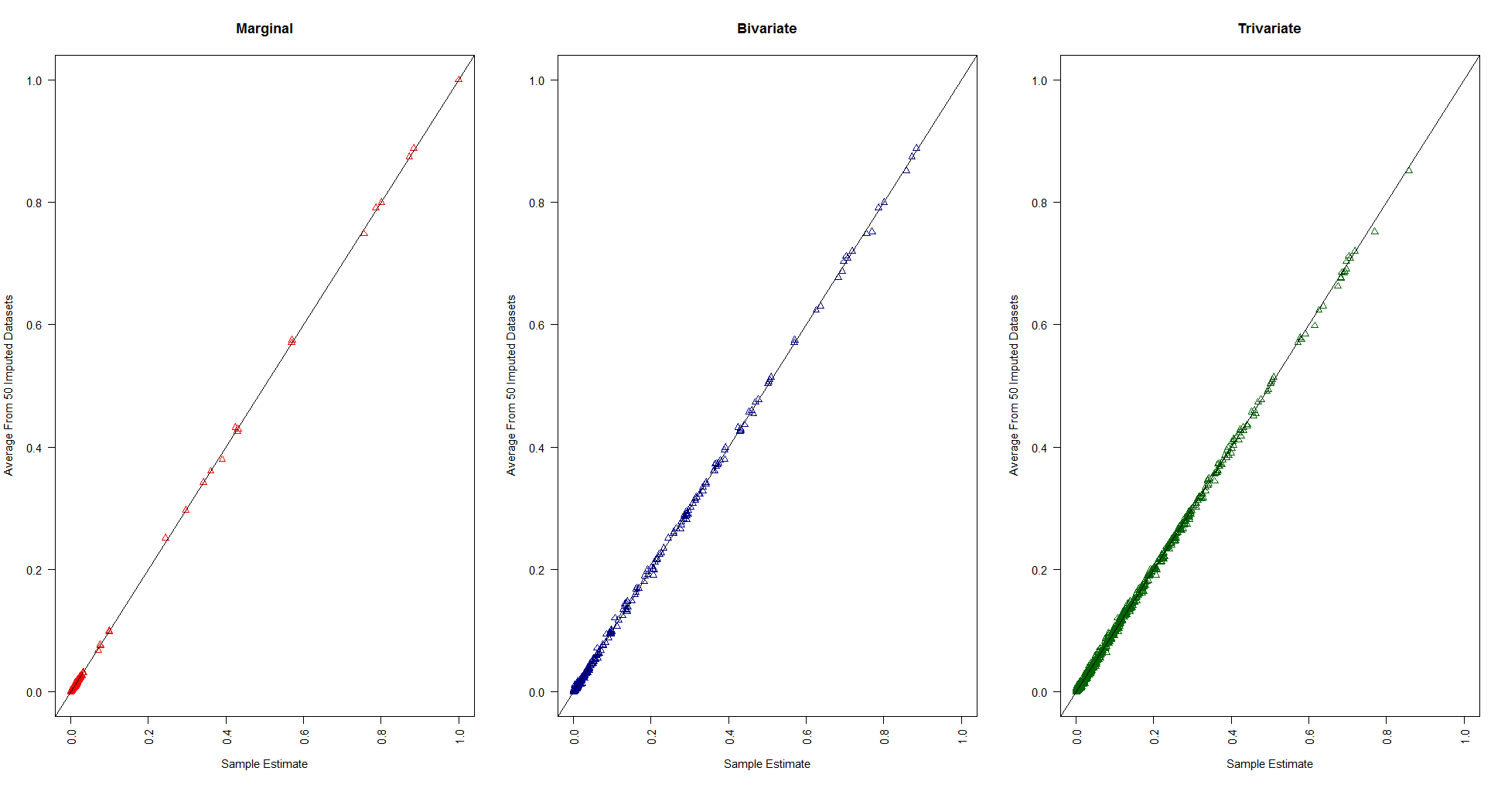}
	\caption{Marginal, bivariate and trivariate probabilities computed in the sample and imputed datasets from the truncated NDPMPM with the rejection sampler.  Household heads' data values moved to the household level.}
	\label{AllProbs}
\end{figure}

\begin{figure}[t]
	\centering
	\includegraphics[width=\textwidth, height = 3.1in, angle=0]{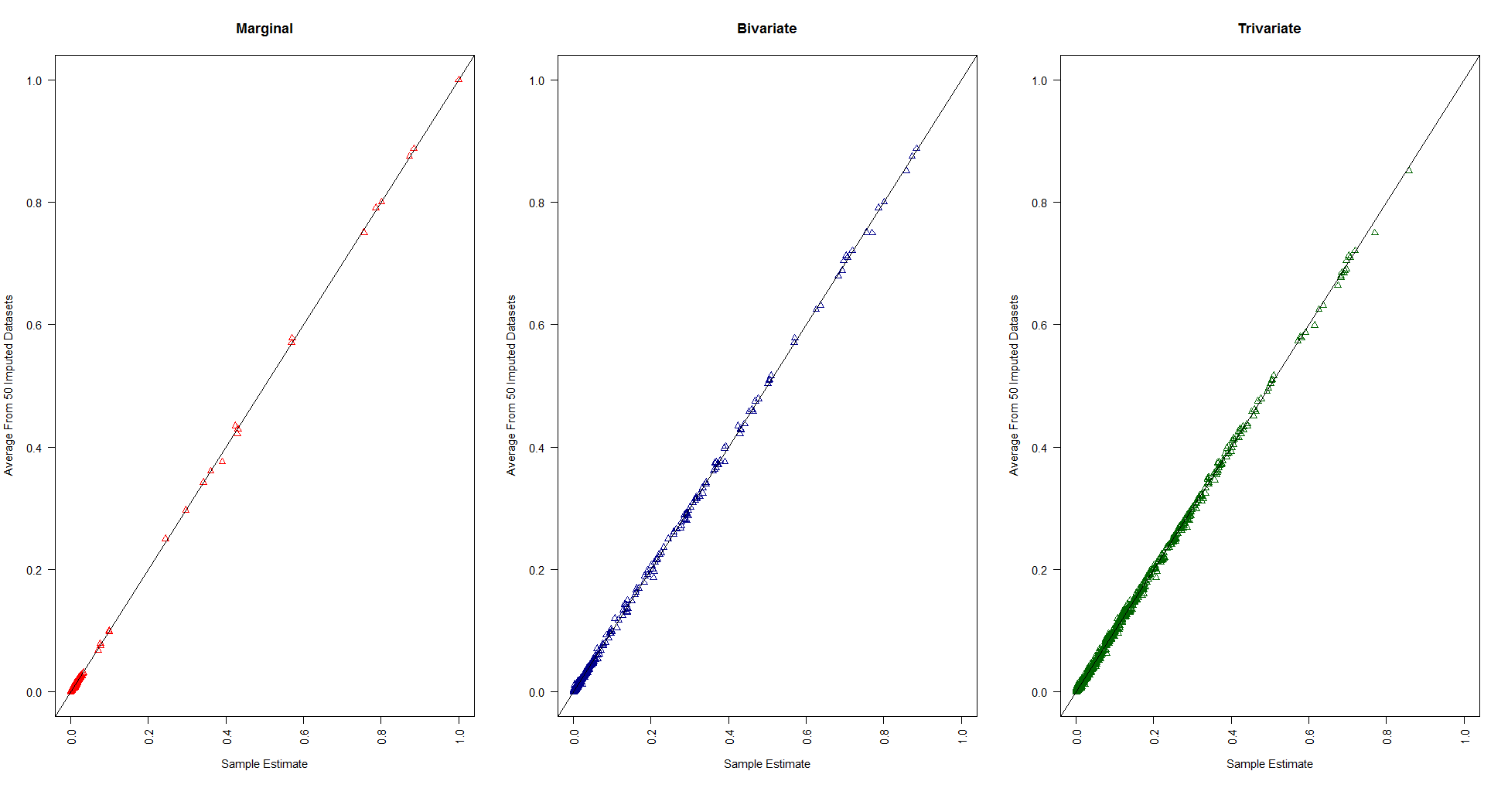}
	\caption{Marginal, bivariate and trivariate probabilities computed in the sample and imputed datasets from the truncated NDPMPM using the cap-and-weight approach. Household heads' data values to the household level.}
	\label{AllProbs_Weighted}
\end{figure}

We estimate the NDPMPM using two approaches, both using the rejection step S9$^\prime$ in Section \ref{MissingData}. 
The first approach considers  $\psi_2 = \psi_3 = \psi_4 = 1$, i.e., without using the cap-and-weight approach, while the second approach considers $\psi_2 = \psi_3 = 1/2$ and $\psi_4 = 1/3$. For each approach, we run the MCMC sampler for 10,000 iterations, discarding the first 5,000 as burn-in and thinning the remaining samples every five iterations, resulting in 1,000 MCMC post burn-in iterates.  We set $F = 30$  and $S = 15$ for each approach based on initial tuning runs.   Across the approaches, the effective number of occupied household-level clusters usually ranges from 13 to 16 with a maximum of 25, while the effective number of occupied individual-level clusters across all household-level clusters ranges from 3 to 5 with a maximum of 10. For convergence, we examined trace plots of $\alpha$, $\beta$, and weighted averages of a random sample of the multinomial probabilities in (\ref{ModelSpecification1}) and (\ref{ModelSpecification2}) (since the multinomial probabilities themselves are prone to label switching).

For both methods, we generate $L = 50$ completed datasets, $\textbf{Z} = (\textbf{Z}^{(1)}, \ldots, \textbf{Z}^{(50)})$, using the posterior predictive distribution of the NDPMPM, from which we estimate all marginal distributions, bivariate distributions of all possible pairs of variables, and trivariate distributions of all possible triplets of variables. We also estimate several probabilities that depend on within household relationships and the household head to investigate the performance of the NDPMPM in estimating complex relationships. We obtain confidence intervals using multiple imputation inferences \citep{Rubin1987}. As a brief review, let $q$ be the completed-data point estimator of some estimand $Q$, and let $u$ be the estimator of variance associated with $q$.  For $l=1, \dots, L$, let $q^{(l)}$ and $u^{(l)}$ be the values of $q$ and $u$ in completed dataset $\textbf{Z}^{(l)}$.  We use $\bar{q}_L = \sum_{l=1}^L q^{(l)}/L$ as the point estimate of $Q$.  We use $T_L = (1 + 1/L)b_L + \bar{u}_L$ as the estimated variance of $\bar{q}$, where $b_L = \sum_{l=1}^L (q^{(l)} - \bar{q}_L)^2/(L-1)$ and $\bar{u}_L = \sum_{l=1}^L u^{(l)}/L$.  We make inference about $Q$ using $(\bar{q}_L - Q) \sim t_{v}(0, T_L)$, where $t_{v}$ is a $t$-distribution with $v = (L-1)(1 + \bar{u}_L/ [(1+1/L) b_L])^2$ degrees of freedom.

Figures \ref{AllProbs} and \ref{AllProbs_Weighted} display the value of $\bar{q}_{50}$ for each estimated marginal, bivariate and trivariate probability plotted against its corresponding estimate from the original data, without missing values. Figure \ref{AllProbs} shows the results for the NDPMPM with the rejection sampler, and Figure \ref{AllProbs_Weighted} shows the results for the NDPMPM using the cap-and-weight approach. For both approaches, the point estimates are close to those from the data before introducing missing values, suggesting that the NDPMPM does a good job of capturing important features of the joint distribution of the variables. Figure \ref{AllProbs_Weighted} in particular also shows that the cap-and-weight approach did not degrade the estimates.

\begin{table}[t!]
	\footnotesize
	\centering
	\caption{Confidence intervals for selected probabilities that depend on within-household relationships in the original and imputed datasets. ``No missing'' is based on the sampled data before introducing missing values, ``NDPMPM'' uses the truncated NDPMPM, moving household heads' data values to the household level, and ``NDPMPM Capped'' uses the truncated NDPMPM with the cap-and-weight approach and moving household heads' data values to the household level. ``HH '' means household head, ``SP'' means spouse, ``CH'' means child, and ``CP'' means couple. $Q$ is the value in the full population of 764,580 households.} \label{CI:MissingData}
	\begin{tabular}[c]{L{0.49\textwidth}|R{0.03\textwidth}R{0.115\textwidth}R{0.115\textwidth}R{0.115\textwidth}}
		& $Q$ & No Missing & NDPMPM & NDPMPM Capped \\ 
		\hline
		All same race household: & & & & \\ 
		\hspace{8pt}  $n_i = 2$ & .942 & (.932, .949) & (.891, .917) & (.884, .911) \\ 
		\hspace{8pt}  $n_i = 3$ & .908 & (.907, .937) & (.843, .890) & (.821, .870) \\ 
		\hspace{8pt}  $n_i = 4$ & .901 & (.879, .917) & (.793, .851) & (.766, .828) \\ 
		SP present & .696 & (.682, .707) & (.695, .722) & (.695, .722)  \\ 
		Same race CP & .656 & (.641, .668) & (.640, .669) & (.634, .664) \\ 
		SP present, HH is White & .600 & (.589, .616) & (.603, .632) & (.604, .634)  \\ 
		White CP & .580 & (.569, .596) & (.577, .606) & (.574, .604) \\ 
		CP with age difference less than five & .488 & (.465, .492) & (.341, .371) & (.324, .355) \\ 
		Male HH, home owner & .476 & (.456, .484) & (.450, .479) & (.451, .480) \\ 
		HH over 35, no CH present & .462 & (.441, .468) & (.442, .470) & (.443, .471) \\ 
		At least one biological CH present & .437 & (.431, .458) & (.430, .459) & (.428, .456) \\ 
		HH older than SP, White HH & .322 & (.309, .335) & (.307, .339) & (.311, .343)  \\ 
		Adult female w/ at least one CH under 5 & .078 & (.070, .085) & (.062, .078) & (.061, .077) \\ 
		White HH with Hisp origin & .066 & (.064, .078) & (.062, .079) & (.062, .078)  \\ 
		Non-White CP, home owner & .058 & (.050, .063) & (.038, .052) & (.037, .051)  \\ 
		Two generations present, Black HH & .057 & (.053, .066) & (.052, .066) & (.052, .067) \\ 
		Black HH, home owner & .052 & (.046, .058) & (.044, .058) & (.044, .059)  \\
		SP present, HH is Black & .039 & (.032, .042) & (.032, .044) & (.031, .043)  \\ 
		White-nonwhite CP & .034 & (.029, .039) & (.038, .053) & (.043, .059) \\ 
		Hisp HH over 50, home owner & .029 & (.025, .034) & (.023, .034) & (.024, .034)\\ 
		One grandchild present & .028 & (.023, .033) & (.024, .035) & (.023, .035)  \\ 
		Adult Black female w/ at least one CH under 18 & .027 & (.028, .038) & (.025, .036) & (.025, .036) \\ 
		At least two generations present, Hisp CP & .027 & (.022, .031) & (.022, .032) & (.023, .033) \\ 
		Hisp CP with at least one biological CH & .025 & (.020, .028) & (.019, .029) & (.020, .030)  \\ 
		At least three generations present & .023 & (.020, .028) & (.017, .026) & (.017, .026)\\ 
		Only one parent & .020 & (.016, .024) & (.013, .021) & (.013, .021)  \\ 
		At least one stepchild & .019 & (.018, .026) & (.019, .030) & (.019, .030)  \\ 
		Adult Hisp male w/ at least one CH under 10 & .018 & (.017, .025) & (.014, .022) & (.014, .022)  \\ 
		At least one adopted CH, White CP & .008 & (.005, .010) & (.004, .010) & (.004, .011) \\ 
		Black CP with at least two biological children & .006 & (.003, .007) & (.003, .007) & (.003, .007)  \\ 
		Black HH under 40, home owner & .005 & (.005, .009) & (.006, .013) & (.007, .013) \\ 
		Three generations present, White CP & .005 & (.004, .008) & (.004, .010) & (.004, .009)  \\ 
		White HH under 25, home owner & .003 & (.002, .005) & (.003, .007) & (.003, .007) \\ 
		
		\hline
	\end{tabular}
\end{table}

Table \ref{CI:MissingData} displays $95\%$ confidence intervals for several probabilities involving within-household relationships, as well as the value in the full population of 764,580 households.  The intervals include the two based on the NDPMPM imputation engines and the interval from the data before introducing missingness. For the latter, we use the usual Wald interval, $\hat{p} \pm 1.96 \sqrt{\hat{p}(1-\hat{p})/n}$, where $\hat{p}$ is the corresponding sample percentage. For the most part, the intervals from the NDPMPM with the full rejection sampling are close to those based on the data without any missingness. They tend to include the true population quantity.  The NDPMPM imputation engine results in noticeable downward bias for the percentages of households where everyone is the same race, with bias increasing as the household size gets bigger. 
This is a challenging estimand to estimate accurately via imputation, particularly for larger households.  
\citet{HuEtAl2018} identified biases in the same direction when using the NDPMPM (with household head data treated as individual-level variables) to generate fully synthetic data, noting that the bias gets smaller as the sample size increases.  The NDPMPM fits the joint distribution of the data better and better as the sample size grows.  Hence, we expect the NDPMPM imputation engine to be more accurate with larger sample sizes, as well as with smaller fractions of missing values.  

The interval estimates from the cap-and-weight method are generally similar to those for the full rejection sampler, with some degradation particularly  for the percentages of same race households by household size. This degradation comes with a benefit, however.
Based on MCMC runs on a standard laptop, the NDPMPM using the cap-and-weight approach and moving household heads' data values to the household level is about $42\%$ faster than the NDPMPM with household heads' data values moved to the household level. 

\section{Discussion} \label{Discussion}

The empirical study suggests that the NDPMPM can provide high quality imputations for categorical data nested within households. To our knowledge, this is the first parametric imputation engine for nested multivariate categorical data.  The study also illustrates that, with modest sample sizes, agencies should not expect the NDPMPM to preserve  all features of the joint distribution. 
Of course, this is the case with any imputation engine.  For the NDPMPM, agencies may be able to improve accuracy for targeted quantities by recoding the data used to fit the model.  For example, one can create a new household-level variable that equals one when  everyone has the same race and equals zero otherwise, and replace the individual race variable with a new variable that has levels ``1 = race is the same as race of household head,'' ``2 = race is white and differs from race of household head,'' ``3 = race is black and differs from race of household head,'' and so on.  The NDPMPM would be estimated with the household-level same race variable and the new individual-level race variable.  This would encourage the NDPMPM to estimate the percentages with the same race very accurately, as it would be just another household-level variable like home ownership.  It also would add structural zeros involving race to the computation.  Evaluating the trade offs in accuracy and computational costs of such recodings is a topic for future research.

The NDPMPM can be computationally expensive, even with the speed-ups presented in this article.  The expensive parts of the algorithm are the rejection sampling steps.
Fortunately, these can be done easily by parallel processing.  For example, we can require each processor to generate a fraction of the impossible cases in Section \ref{NDPMPMSampler}.  We also can spread the rejection steps for the imputations over many processors.  These steps should cut run time by a factor roughly equal to the  number of processors available.

The empirical study used households up to size four.  We have run the model on data with households up to size seven in reasonable time (a few hours on a standard laptop). Accuracy results are similar qualitatively.  As the household sizes get large, the model can generate hundreds or even thousands times as many impossible households as there are feasible ones, slowing the algorithm.  In such cases, the cap-and-weight approach is essential for practical applications. 

\section{Acknowledgments}

This research was supported by grants from the National Science Foundation (NSF SES 1131897) and the  Alfred P. Sloan Foundation (G-2-15-20166003).

\newpage
\section{Supplementary Materials}
This is a supplementary material to the paper. It contains proof that the rejection sampling step S9$^\prime$ in Section \ref{MissingData} generates samples from the correct posterior distribution. It also contains the modified Gibbs sampler for the cap-and-weight approach and a list of the structural zero rules used in fitting the NDPMPM model. Finally, we include empirical results for the speedup approaches mentioned in the paper, using synthetic data, and additional results for handling missing data using the NDPMPM under a missing completely at random scenario.

\subsection{Proof that the rejection sampling step S9$^\prime$ in Section \ref{MissingData} generates samples from the correct posterior distribution}

The $X^1_{ik}$ and $X^1_{ijk}$ values generated using the rejection sampler in Step S9$^\prime$ are generated from the full conditionals, resulting in a valid Gibbs sampler.  The proof follows from the properties of rejection sampling (or simple accept reject). The target distribution is the full conditional for $\textbf{X}_i^{\textrm{mis}}$.  It can be re-expressed as
$$p(\textbf{X}_i^{\textrm{mis}}) \ = \dfrac{\mathds{1}\{\textbf{X}_i^1 \notin \mathcal{S}_h \}}{\Pr(\textbf{X}_i \notin \mathcal{S}_h | \theta)}   g(\textbf{X}_i^{\textrm{mis}})$$
where 
$$g(\textbf{X}_i^{\textrm{mis}})  = \pi_{G^1_i}\prod^{p+q}_{k | a_{ik} = 1} \lambda^{(k)}_{G^1_iX^1_{ik}} \left(\prod^{n_i}_{j=1} \omega_{G^1_iM^1_{ij}}\prod^p_{k | b_{ijk} = 1} \phi^{(k)}_{G^1_iM^1_{ij}X^1_{ijk}} \right).$$
Our rejection scheme uses $g(\textbf{X}_i^{\textrm{mis}})$ as a proposal for $p(\textbf{X}_i^{\textrm{mis}})$. To show that the draws are indeed from $p(\textbf{X}_i^{\textrm{mis}})$, we need to verify that $w(\textbf{X}_i^{\textrm{mis}}) = p(\textbf{X}_i^{\textrm{mis}})/g(\textbf{X}_i^{\textrm{mis}}) < M$, where $1 < M < \infty$, and that we are accepting each sample with probability $w(\textbf{X}_i^{\textrm{mis}})/M$. In our case, 

\begin{enumerate}
	\item $w(\textbf{X}_i^{\textrm{mis}}) = p(\textbf{X}_i^{\textrm{mis}})/g(\textbf{X}_i^{\textrm{mis}}) = \mathds{1}\{\textbf{X}_i^1 \notin \mathcal{S}_h \}/\Pr(\textbf{X}_i \notin \mathcal{S}_h | \theta)  \leq  1/\Pr(\textbf{X}_i \notin \mathcal{S}_h | \theta)$, and $0 < \Pr(\textbf{X}_i \notin \mathcal{S}_h | \theta) < 1 \ \Rightarrow \ 1 < 1/\Pr(\textbf{X}_i \notin \mathcal{S}_h | \theta) < \infty$ necessarily.
	
	\item By sampling until we obtain a valid sample that satisfies $\textbf{X}_i^1 \notin \mathcal{S}_h$, we are indeed sampling with probability $w(\textbf{X}_i^{\textrm{mis}})/M = \mathds{1}\{\textbf{X}_i^1 \notin \mathcal{S}_h \}$.
\end{enumerate}

\subsection{Modified Gibbs sampler for the cap-and-weight approach}

The modified Gibbs sampler for the cap-and-weight approach replaces steps S1, S3, S4, S5 and S6 of the Gibbs sampler in the main text as follows.
\begin{enumerate}
	\item[S1$^\star$.] For each $h \in \mathcal{H}$, repeat steps S1(a) to S1(e) as before but modify step S1(f) to: if $t_1 < \lceil n_{1h} \times \psi_h \rceil$, return to step (b). Otherwise, set $n_{0h} = t_0$. 
	
	\item[S3$^\star$.] Set $u_F = 1$. Sample	
	\begin{equation*}
	\begin{split}
	u_g | - \ & \sim \textrm{Beta} \left(1 + U_g, \alpha + \sum^F_{f=g+1} U_f  \right), \ \ \pi_g = u_g \prod_{f<g} (1 - u_f) \\
	\textrm{where} \ \ U_g & = \sum^{n}_{i=1} \mathds{1}(G^1_i = g) + \sum_{h \in \mathcal{H}} \dfrac{1}{\psi_h} \sum\limits_{i | n^0_i = h} \mathds{1}(G_i^0 = g)
	\end{split}
	\end{equation*}
	for $g = 1, \ldots, F-1$.
	
	\item[S4$^\star$.] Set $v_{gM} = 1$ for for $g = 1, \ldots, F$. Sample
	\begin{equation*}
	\begin{split}
	v_{gm} | - \ & \sim \textrm{Beta} \left(1 + V_{gm}, \beta + \sum^S_{s=m+1} V_{gs}  \right), \ \ \omega_{gm} = v_{gm} \prod_{s<m} (1 - v_{gs}) \\
	\textrm{where} \ \ V_{gm} & = \sum^{n}_{i=1} \mathds{1}(M^1_{ij} = m, G^1_i = g) + \sum_{h \in \mathcal{H}} \dfrac{1}{\psi_h} \sum\limits_{i | n_i^0 = h} \mathds{1}(M_{ij}^0 = m, G_i^0 = g) 
	\end{split}
	\end{equation*}
	for $m = 1, \ldots, S-1$ and $g = 1, \ldots, F$.
	
	\item[S5$^\star$.] Sample 
	\begin{equation*}
	\begin{split}
	\lambda_g^{(k)} | - & \sim \textrm{Dirichlet}\left(1 + \eta^{(k)}_{g1}, \ldots, 1 + \eta^{(k)}_{gd_k} \right) \\
	\textrm{where} \ \ \eta^{(k)}_{gc} & = \sum^{n}_{i|G^1_i = g} \mathds{1}(X^1_{ik} = c) + \sum_{h \in \mathcal{H}} \dfrac{1}{\psi_h} \sum\limits_{i \big| \substack{n_i^0 = h, \\ G_i^0 = g}} \mathds{1}(X_{ik}^0 = c)
	\end{split}
	\end{equation*}
	for $g = 1, \ldots, F$ and $k = p+1, \ldots, q$. 
	
	\item[S6$^\star$.] Sample 
	\begin{equation*}
	\begin{split}
	\phi_{gm}^{(k)} | - & \sim \textrm{Dirichlet}\left(1 + \nu^{(k)}_{gm1}, \ldots, 1 + \nu^{(k)}_{gmd_k} \right) \\
	\textrm{where} \ \ \nu^{(k)}_{gmc} & = \sum^{n}_{i \big| \substack{G^1_i = g, \\  M^1_{ij} = m}} \mathds{1}(X_{ijk}^1 = c) + \sum_{h \in \mathcal{H}} \dfrac{1}{\psi_h} \sum\limits_{i \big| \substack{n_i^0 = h, \\ G_i^0 = g, \\  M^0_{ij} = m}} \mathds{1}(X_{ijk}^0 = c)
	\end{split}
	\end{equation*}
	for $g = 1, \ldots, F$, $m = 1, \ldots, S$ and $k = 1, \ldots, p$.
\end{enumerate}

\subsection{List of structural zeros}
We fit the NDPMPM model using structural zeros which involve ages and relationships of individuals in the same house. The full list of the rules used is presented in Table \ref{SynStructuralZeros}. These rules were derived from the 2012 ACS by identifying combinations involving the relationship variable that do not appear in the constructed population. This list should not be interpreted as a ``true'' list of impossible combinations in census data.

\begin{table}[h!]
	\small
	\centering
	\caption{List of structural zeros.} \label{SynStructuralZeros}
	\begin{tabular}[c]{p{0.03\textwidth}p{0.9\textwidth}}
		\hline  \hline 
		& Description \\ \hline 
		& \\ 
		& \underline{Rules common to generating both the synthetic and imputed datasets} \\
		1. & Each household must contain exactly one head and he/she must be at least 16 years old. \\
		2. & Each household cannot contain more than one spouse and he/she must be at least 16 years old. \\ 
		3. & Married couples are of opposite sex, and age difference between individuals in the couples cannot exceed 49.\\
		4. & The youngest parent must be older than the household head by at least 4. \\
		5. & The youngest parent-in-law must be older than the household head by at least 4. \\
		6. & The age difference between the household head and siblings cannot exceed 37. \\
		7. & The household head must be at least 31 years old to be a grandparent and his/her spouse must be at least 17. Also, He/she must be older than the oldest grandchild by at least 26. \\
		
		& \\ 
		& \underline{Rules specific to generating the synthetic datasets} \\
		8. & The household head must be older than the oldest child by at least 7. \\
		& \\
		& \underline{Rules specific to generating the imputed datasets} \\
		9. & The household head must be older than the oldest biological child by at least 7. \\
		10. & The household head must be older than the oldest adopted child by at least 11. \\
		11. & The household head must be older than the oldest stepchild by at least 9. \\
		\hline
	\end{tabular}
\end{table}

\subsection{Empirical study of the speedup approaches} \label{SpeedUpResults}

We evaluate the performance of the two speedup approaches mentioned in the main text using synthetic data. We use data from the public use microdata files from the 2012 ACS, available for download from the United States Census Bureau (\url{http://www2.census.gov/acs2012_1yr/pums/}) to construct a population of $857,018$ households of sizes $\mathcal{H} = \{2, 3, 4, 5, 6\}$, from which we sample $n=10,000$ households comprising  $N=29,117$ individuals. We work with the variables described in Table \ref{NDPMPMVariableDef}.  We evaluate the approaches using probabilities that depend on within household relationships and the household head.

\begin{table}[t!]
	\small
	\centering
	\caption{Description of variables used in the synthetic data illustration} \label{NDPMPMVariableDef}
	\begin{tabular}[c]{p{0.4\textwidth}p{0.5\textwidth}}
		Description of variable & Categories \\ \hline 
		& \\
		\multicolumn{2}{l}{\underline{Household-level variables}} \\
		Ownership of dwelling & 1 = owned or being bought, 2 = rented\\
		Household size & 2 = 2 people, 3 = 3 people, 4 = 4 people, \\
		& 5 = 5 people, 6 = 6 people \\
		& \\
		\multicolumn{2}{l}{\underline{Individual-level variables}} \\
		
		Gender & 1 = male, 2 = female \\
		Race & 1 = white, 2 = black,\\
		& 3 = American Indian or Alaska native, \\
		& 4 = Chinese, 5 = Japanese, \\
		& 6 = other Asian/Pacific islander, 7 = other race, \\
		& 8 = two major races, \\
		& 9 = three or more major races \\
		Hispanic origin & 1 = not Hispanic, 2 = Mexican, \\
		& 3 = Puerto Rican, 4 = Cuban, 5 = other \\
		Age & 1 = less than one year old, 2 = 1 year old,\\
		& 3 = 2 years old, \ldots, 96 = 95 years old \\
		Relationship to head of household & 1 = household head, 2 = spouse, 3 = child, \\
		& 4 = child-in-law, 5 = parent, 6 = parent-in-law,\\
		& 7 = sibling, 8 = sibling-in-law, 9 = grandchild, \\
		& 10 = other relative, 11 = partner/friend/visitor, \\
		& 12 = other non-relative \\ \hline
	\end{tabular}
\end{table}

We consider the NDPMPM using two approaches, both moving the values of the household head to the household level as in Section \ref{SpeedUpMoveHH} of the main text and also using the cap-and-weight approach in Section \ref{SpeedUpCapping} of the main text. The first approach considers  $\psi_2 = \psi_3 = \psi_4 = \psi_5 = \psi_6 = 1$ while the second approach considers $\psi_2 = \psi_3 = 1/2$ and $\psi_4 = \psi_5 = \psi_6 = 1/3$. We compare these approaches to the NDPMPM as presented in \cite{HuEtAl2018}. For each approach, we create $L = 50$ synthetic datasets, $\textbf{Z} = (\textbf{Z}^{(1)}, \ldots, \textbf{Z}^{(50)})$.
We generate the synthetic datasets so that the number of households of size $h \in \mathcal{H}$ in each $\textbf{Z}^{(l)}$ exactly matches $n_h$ from the observed data. 
Thus, $\textbf{Z}$ comprises partially synthetic data \citep{Little1993, Reiter2003}, even though every released $Z_{ijk}$ is a simulated value. We combine the estimates using using the approach in \citet{Reiter2003}. As a brief review, let $q$ be the point estimator of some estimand $Q$, and let $u$ be the estimator of variance associated with $q$.  For $l=1, \dots, L$, let $q_l$ and $u_l$ be the values of $q$ and $u$ in synthetic dataset $\mathbf{Z}^{(l)}$.  We use $\bar{q} = \sum_{l=1}^L q_l/L$ as the point estimate of $Q$ and $T = \bar{u} + b/L$ as the estimated variance of $\bar{q}$, where $b = \sum_{l=1}^L (q_l - \bar{q})^2/(L-1)$ and $\bar{u} = \sum_{l=1}^L u_l/L$.  We make inference about $Q$ using $(\bar{q} - Q) \sim t_{v}(0, T)$, where $t_{v}$ is a $t$-distribution with $v = (L-1)(1 + L\bar{u}/b])^2$ degrees of freedom.

\begin{table}[t!]
	\small
	\centering
	\caption{Confidence intervals for selected probabilities that depend on within-household relationships in the original and synthetic datasets. ``Original'' is based on the sampled data, ``NDPMPM'' is the default MCMC sampler described in Section 2.2 of the main text, ``NDPMPM w/ HH moved'' is the default sampler, moving household heads' data values to the household level, ``NDPMPM capped w/ HH moved'' uses the cap-and-weight approach and moving household heads' data values to the household level. ``HH '' means household head and ``SP'' means spouse.}
	\label{CI:SyntheticData}
	\begin{tabular}[c]{L{0.3\textwidth}|R{0.14\textwidth}R{0.14\textwidth}R{0.14\textwidth}R{0.14\textwidth}}
		& Original & NDPMPM & NDPMPM w/ HH moved & NDPMPM capped w/ HH moved \\ 
		\hline
		All same race & & & & \\
		\hspace{6pt} $n_i = 2$ & (.939, .951) & (.918, .932) & (.912, .928) & (.910, .925) \\ 
		\hspace{6pt} $n_i = 3$ & (.896, .920) & (.859, .888) & (.845, .875) & (.844, .874) \\ 
		\hspace{6pt} $n_i = 4$ & (.885, .912) & (.826, .860) & (.813, .848) & (.817, .852) \\ 
		\hspace{6pt} $n_i = 5$ & (.879, .922) & (.786, .841) & (.786, .841) & (.777, .834) \\ 
		\hspace{6pt} $n_i = 6$ & (.831, .910) & (.701, .803) & (.718, .819) & (.660, .768) \\ 
		SP present & (.693, .711) & (.678, .697) & (.676, .695) & (.677, .695) \\ 
		SP with white HH & (.589, .608) & (.577, .597) & (.576, .595) & (.575, .595) \\ 
		SP with black HH & (.036, .043) & (.035, .043) & (.034, .042) & (.034, .042) \\ 
		White couple & (.570, .589) & (.560, .580) & (.553, .573) & (.552, .572) \\ 
		White couple, own & (.495, .514) & (.468, .488) & (.461, .481) & (.463, .483) \\ 
		Same race couple & (.655, .673) & (.636, .655) & (.626, .645) & (.625, .644) \\ 
		White-nonwhite couple & (.028, .035) & (.028, .035) & (.034, .041) & (.036, .044) \\ 
		Nonwhite couple, own & (.057, .067) & (.047, .056) & (.045, .053) & (.045, .054) \\ 
		Only mother present & (.017, .022) & (.014, .019) & (.014, .019) & (.013, .018) \\ 
		Only one parent present & (.021, .026) & (.026, .032) & (.026, .033) & (.027, .033) \\ 
		Children present & (.507, .527) & (.493, .512) & (.517, .537) & (.511, .531) \\ 
		Siblings present & (.022, .028) & (.027, .034) & (.027, .033) & (.027, .033) \\ 
		Grandchild present & (.041, .049) & (.051, .060) & (.049, .058) & (.050, .059) \\ 
		Three generations present & (.036, .044) & (.037, .045) & (.042, .050) & (.040, .048) \\ 
		White HH, older than SP & (.309, .327) & (.283, .301) & (.294, .313) & (.302, .321) \\ 
		Nonhisp HH & (.882, .894) & (.875, .888) & (.879, .891) & (.876, .889) \\ 
		White, Hisp HH & (.071, .082) & (.074, .085) & (.072, .082) & (.073, .084) \\ 
		Same age couple & (.087, .098) & (.027, .034) & (.023, .029) & (.024, .031) \\ 
		\hline
	\end{tabular}
\end{table}

For each approach, we run the MCMC sampler for 20,000 iterations, discarding the first 10,000 as burn-in and thinning the remaining samples every five iterations, resulting in 2,000 MCMC post burn-in iterates. We create the $L = 50$ synthetic datasets by randomly sampling from the 2,000 iterates. We set $F = 40$  and $S = 15$ for each approach based on initial tuning runs. For convergence, we examined trace plots of $\alpha$, $\beta$ and weighted averages of a random sample of the multinomial probabilities in the NDPMPM likelihood. Across the approaches, the effective number of occupied household-level clusters usually ranges from 20 to 33 with a maximum of 38, while the effective number of occupied individual-level clusters across all household-level clusters ranges from 5 to 9 with a maximum of 12.

Based on MCMC runs on a standard laptop, moving household heads' data values to the household level alone results in a speedup of about $63\%$ on the default rejection sampler while the cap-and-weight approach alone results in a speedup of about $40\%$.

Table \ref{CI:SyntheticData} shows the $95\%$ confidence intervals for each approach. Essentially, all three approaches result in similar confidence intervals, suggesting not much loss in accuracy from the speedups. Most intervals also are reasonably similar to confidence intervals  based on the original data, except for the percentage of same age couples. The last row is a rigorous test of how well each method can estimate a probability that can be fairly difficult to estimate accurately. In this case, the probability that a household head and spouse are the same age can be difficult to estimate since each individual's age can take 96 different values. All three approaches are thus off from the estimate from the original data in this case. 
These results suggest that we can significantly speedup the sampler with minimal loss in accuracy of estimates and confidence intervals of population estimands.

\subsection{Empirical study of missing data imputation under MCAR} \label{MissingDataResults}

\begin{figure}[t!]
	\centering
	\includegraphics[width=\textwidth, height = 3.1in, angle=0]{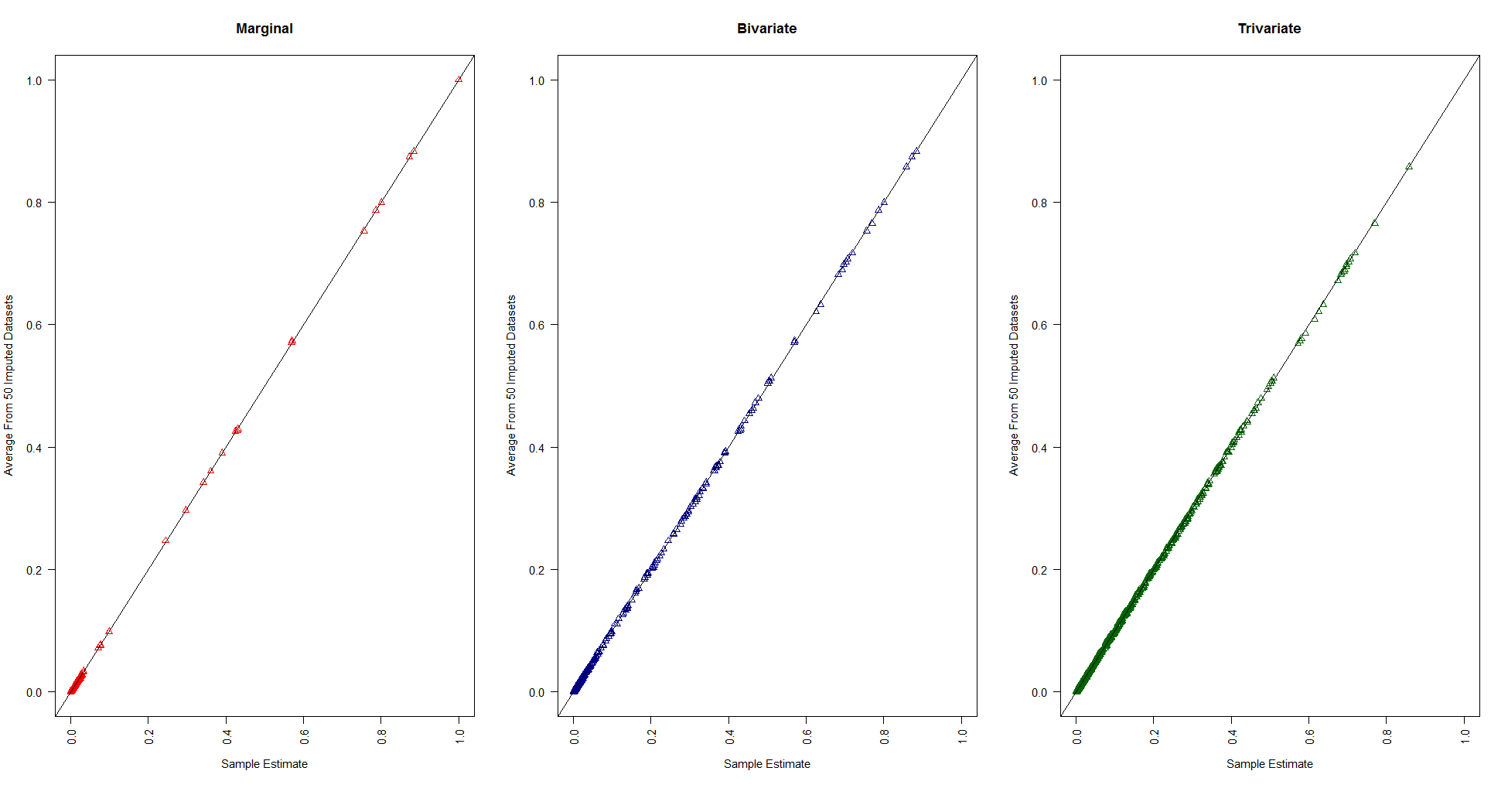}
	\caption{Marginal, bivariate and trivariate probabilities computed in the sample and imputed datasets under MCAR from the truncated NDPMPM with the rejection sampler.  Household heads' data values moved to the household level.}
	\label{AllProbsII}
\end{figure}

We also evaluate the performance of the NDPMPM as an imputation method under a missing completely at random (MCAR) scenario. We use the same data as in Section 5 of the main text. As a reminder, the data contains $n=5,000$ households of sizes $\mathcal{H} = \{2, 3, 4\}$, comprising $N=13,181$ individuals. We introduce missing values using a MCAR scenario. We randomly select 80\% households to be complete cases for all variables. For the remaining 20\%, we let the variable ``household size'' be fully observed and randomly -- and independently -- blank 50\% of each variable for the remaining household-level and individual-level variables. We use these low rates to mimic the actual rates of item nonresponse in census data. 

\begin{figure}[t!]
	\centering
	\includegraphics[width=\textwidth, height = 3.1in, angle=0]{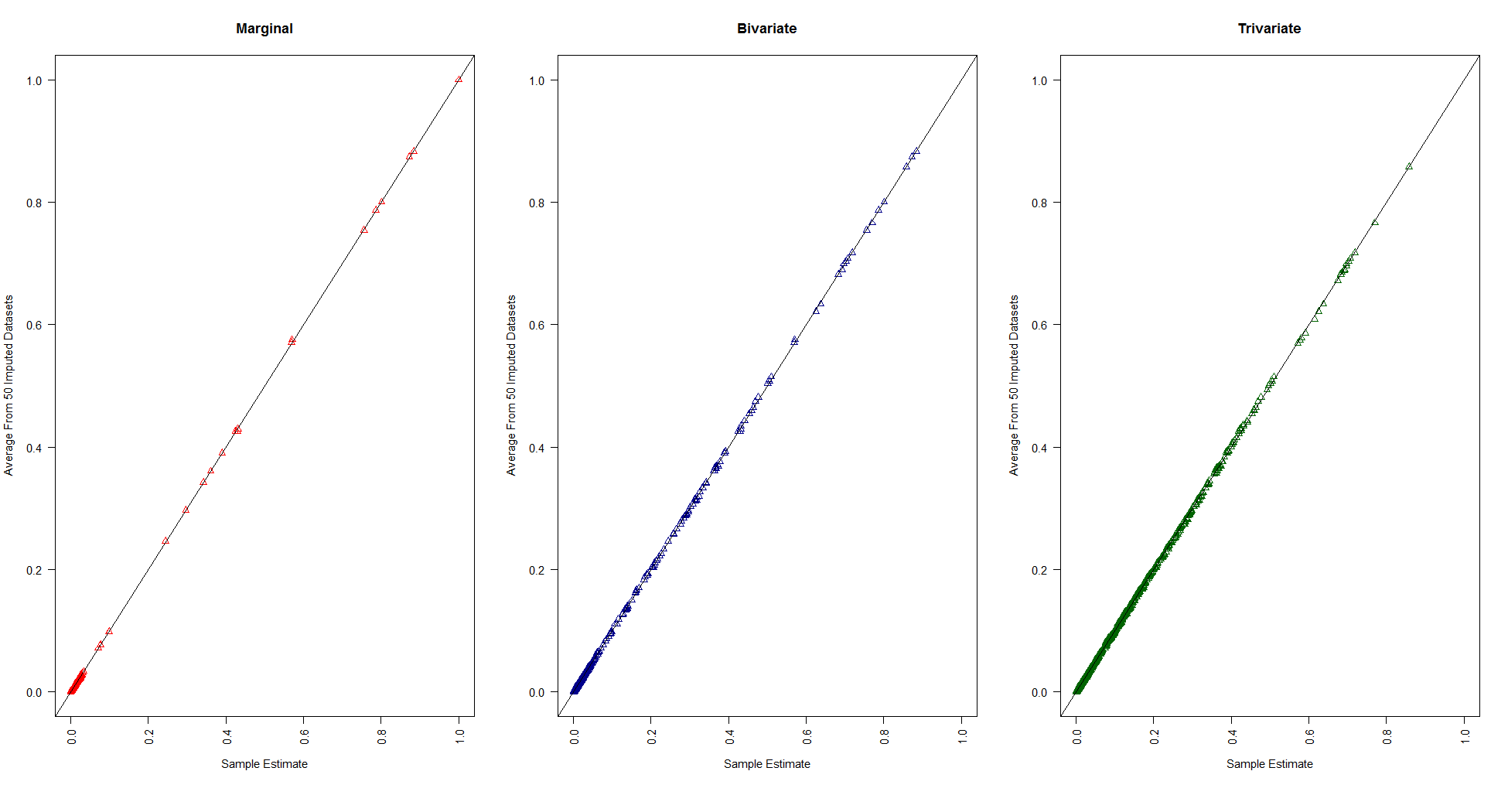}
	\caption{Marginal, bivariate and trivariate probabilities computed in the sample and imputed datasets under MCAR from the truncated NDPMPM using the cap-and-weight approach. Household heads' data values to the household level.}
	\label{AllProbs_WeightedII}
\end{figure}

\begin{table}[t!]
	\footnotesize
	\centering
	\caption{Confidence intervals for selected probabilities that depend on within-household relationships in the original and imputed datasets under MCAR. ``No missing'' is based on the sampled data before introducing missing values, ``NDPMPM'' uses the truncated NDPMPM, moving household heads' data values to the household level, and ``NDPMPM Capped'' uses the truncated NDPMPM with the cap-and-weight approach and moving household heads' data values to the household level. ``HH '' means household head, ``SP'' means spouse, ``CH'' means child, and ``CP'' means couple. $Q$ is the value in the full population of 764,580 households.} \label{CI:MissingDataII}
	\begin{tabular}[c]{L{0.49\textwidth}|R{0.03\textwidth}R{0.115\textwidth}R{0.115\textwidth}R{0.115\textwidth}}
		& $Q$ & No Missing & NDPMPM & NDPMPM Capped \\ 
		\hline
		All same race household: & & & & \\ 
		\hspace{8pt}  $n_i = 2$ & .942 & (.932, .949) & (.924, .944) & (.925, .946) \\ 
		\hspace{8pt}  $n_i = 3$ & .908 & (.907, .937) & (.887, .924) & (.890, .925) \\ 
		\hspace{8pt}  $n_i = 4$ & .901 & (.879, .917) & (.854, .900) & (.855, .900) \\ 
		SP present & .696 & (.682, .707) & (.683, .709) & (.683, .709)  \\ 
		Same race CP & .656 & (.641, .668) & (.637, .664) & (.638, .665) \\ 
		SP present, HH is White & .600 & (.589, .616) & (.590, .618) & (.590, .618)  \\ 
		White CP & .580 & (.569, .596) & (.568, .596) & (.568, .597) \\ 
		CP with age difference less than five & .488 & (.465, .492) & (.422, .451) & (.422, .450) \\ 
		Male HH, home owner & .476 & (.456, .484) & (.455, .483) & (.456, .485) \\ 
		HH over 35, no CH present & .462 & (.441, .468) & (.438, .466) & (.438, .466) \\ 
		At least one biological CH present & .437 & (.431, .458) & (.432, .460) & (.432, .460) \\ 
		HH older than SP, White HH & .322 & (.309, .335) & (.308, .335) & (.306, .333)  \\ 
		Adult female w/ at least one CH under 5 & .078 & (.070, .085) & (.068, .084) & (.067, .083) \\ 
		White HH with Hisp origin & .066 & (.064, .078) & (.064, .079) & (.064, .079)  \\ 
		Non-White CP, home owner & .058 & (.050, .063) & (.048, .061) & (.048, .061)  \\ 
		Two generations present, Black HH & .057 & (.053, .066) & (.053, .066) & (.053, .067) \\ 
		Black HH, home owner & .052 & (.046, .058) & (.046, .059) & (.046, .059)  \\
		SP present, HH is Black & .039 & (.032, .042) & (.032, .043) & (.032, .042)  \\ 
		White-nonwhite CP & .034 & (.029, .039) & (.032, .044) & (.032, .044) \\ 
		Hisp HH over 50, home owner & .029 & (.025, .034) & (.025, .035) & (.025, .035)\\ 
		One grandchild present & .028 & (.023, .033) & (.024, .034) & (.024, .034)  \\ 
		Adult Black female w/ at least one CH under 18 & .027 & (.028, .038) & (.027, .037) & (.027, .037) \\ 
		At least two generations present, Hisp CP & .027 & (.022, .031) & (.022, .031) & (.022, .031) \\ 
		Hisp CP with at least one biological CH & .025 & (.020, .028) & (.019, .028) & (.019, .028)  \\ 
		At least three generations present & .023 & (.020, .028) & (.019, .028) & (.019, .028)\\ 
		Only one parent & .020 & (.016, .024) & (.016, .024) & (.016, .024)  \\ 
		At least one stepchild & .019 & (.018, .026) & (.018, .027) & (.018, .027)  \\ 
		Adult Hisp male w/ at least one CH under 10 & .018 & (.017, .025) & (.016, .025) & (.016, .025)  \\ 
		At least one adopted CH, White CP & .008 & (.005, .010) & (.005, .010) & (.005, .010) \\ 
		Black CP with at least two biological children & .006 & (.003, .007) & (.003, .007) & (.003, .007)  \\ 
		Black HH under 40, home owner & .005 & (.005, .009) & (.005, .010) & (.005, .011) \\ 
		Three generations present, White CP & .005 & (.004, .008) & (.004, .010) & (.004, .009)  \\ 
		White HH under 25, home owner & .003 & (.002, .005) & (.004, .009) & (.004, .009) \\ 
		
		\hline
	\end{tabular}
\end{table}

Similar to the main text, we estimate the NDPMPM using two approaches, both combining the rejection step in Section 4.1 of the main text with the cap-and-weight approach in Section 4.2 of the main text. The first approach considers  $\psi_2 = \psi_3 = \psi_4 = 1$ while the second approach considers $\psi_2 = \psi_3 = 1/2$ and $\psi_4 = 1/3$. For each approach, we run the MCMC sampler for 10,000 iterations, discarding the first 5,000 as burn-in and thinning the remaining samples every five iterations, resulting in 1,000 MCMC post burn-in iterates.  We set $F = 30$  and $S = 15$ for each approach based on initial tuning runs. 
We monitor convergence as in the main text. For both methods, we generate $L = 50$ completed datasets, $\textbf{Z} = (\textbf{Z}^{(1)}, \ldots, \textbf{Z}^{(50)})$, using the posterior predictive distribution of the NDPMPM, from which we estimate the same probabilities as in the main text.

Figures \ref{AllProbsII} and \ref{AllProbs_WeightedII} display each estimated marginal, bivariate and trivariate probability $\bar{q}_{50}$ plotted against its corresponding estimate from the original data, without missing values. Figure \ref{AllProbsII} shows the results for the NDPMPM with the rejection sampler, and Figure \ref{AllProbs_WeightedII} shows the results for the NDPMPM using the cap-and-weight approach. For both approaches, the NDPMPM does a good job of capturing important features of the joint distribution of the variables as the point estimates are very close to those from the data before introducing missing values. In short, the results are very similar to those in the main text, though more accurate.

Table \ref{CI:MissingDataII} displays $95\%$ confidence intervals for selected probabilities involving within-household relationships, as well as the value in the full population of 764,580 households.  The intervals include the two based on the NDPMPM imputation engines and the interval from the data before introducing missingness. The intervals are generally more accurate than those presented in the main text. This is expected since we use lower rates of missingness in the MCAR scenario. For the most part, the intervals from the NDPMPM with the two approaches tend to include the true population quantity. Again, the NDPMPM imputation engine results in downward bias for the percentages of households where everyone is the same race. As mentioned in the main text, this is a challenging estimand to estimate accurately via imputation, particularly for larger households.

\renewcommand{\refname}{References}
\printbibliography
\end{document}